\documentclass[12pt]{article}
\usepackage[top=1in,bottom=1in, left=1in, right=1in]{geometry}
\usepackage{atbegshi,cite}
\usepackage{amsmath,amssymb,amsbsy,amstext, amsthm, simplewick}
\usepackage{hyperref}
\usepackage{graphicx}
\usepackage{amsfonts}
\usepackage[small]{caption}
\usepackage{upgreek}
\usepackage[titletoc]{appendix}
\usepackage{setspace}
\usepackage{color}
\usepackage{tikz}
\usetikzlibrary{decorations.markings,decorations.pathmorphing,patterns}
\usepackage{pbox}

\def\beq{\begin{eqnarray}}
\def\eeq{\end{eqnarray}}

\newcommand{\newc}{\newcommand}
\newc{\fpi}{f_{\pi}}
\newc{\etap}{\eta^{\prime}}
\newc{\llll}{\langle\lambda\lambda\rangle}
\newc{\FFd}{F^a\tilde F^a}
\newc{\qbar}{{\overline q}}
\newc{\TR}{{\rm Tr}}
\newc{\Kahler}{K\"ahler }
\newc{\Zbb}{{\mathbb Z}}
\newc{\Rt}{{\mathbb R}^3}
\newc{\Rf}{{\mathbb R}^4}
\newc{\So}{{\mathbb S}^1}
\newc{\zt}{{\mathbb Z}_2}
\newc{\RtSo}{{\mathbb R}^3\times{\mathbb S}^1}
\newc{\scriminus}{{\cal I}^-}
\newc{\scriplus}{{\cal I}^+}
\newc{\mpl}{M_p}
\newc{\Ricci}{\mathcal{R}}
\newc{\bv}{\phi}
\newc{\calU}{{\cal U}}
\newc{\calK}{K}
\newc{\calUi}{{\cal U}^{-1}}
\newc{\calG}{{\cal G}}
\newc{\calO}{{\cal O}}
\newc{\calOb}{{\cal O}^\dagger}
\newc{\hphi}{{\hat\phi}}
\newc{\lag}{\mathcal{L}}
\newc{\op}{\mathcal{O}}
\newc{\bo}{\mathrm{O}}
\newc{\Tr}{\mathrm{Tr}}
\renewcommand{\Im}{\mathrm{Im}}
\newc{\abs}[1]{| #1 |}

\newc{\dws}[1]{\textbf{\textcolor{purple}{DS: #1}}}
\newc{\jordy}[1]{\textcolor{red}{JdV: #1}}

\begin{document}
\begin{titlepage}
\begin{flushright}
{\large 
ACFI-T18-16\\
}
\end{flushright}

\vskip 2.2cm

\begin{center}

{\large \bf Indirect Signs of the Peccei-Quinn Mechanism}

\vskip 1.4cm

{Jordy de Vries$^{(a,b)}$, Patrick Draper$^{(a,c)}$, Kaori Fuyuto$^{(a)}$, Jonathan Kozaczuk$^{(a,c)}$, \\ and Dave Sutherland$^{(d)}$ }
\\
\vskip 1cm
{\small $^{(a)}$ Amherst Center for Fundamental Interactions, Department of Physics,\\ University of Massachusetts, Amherst, MA 01003\\
$^{(b)}$ RIKEN BNL Research Center, Brookhaven National Laboratory, Upton, NY 11973-5000\\
$^{(c)}$ Department of Physics, University of Illinois, Urbana, IL 61801\\
$^{(d)}$ Department of Physics, University of California, Santa Barbara, CA 93106
}\\
\vspace{0.3cm}
\vskip 4pt

\vskip 1.5cm

\begin{abstract}
In the Standard Model, the renormalization of the QCD vacuum angle $\theta$ is extremely tiny, and small $\theta$ is technically natural. In the general Standard Model effective field theory (SMEFT), however, $\Delta\theta$ is quadratically divergent, reflecting the fact that new sources of hadronic CP-violation typically produce $\calO(1)$ threshold corrections to $\theta$. The observation of such CP-violating interactions would therefore be in tension with solutions to the strong CP problem in which $\theta=0$ is an ultraviolet boundary condition, 
pointing to the Peccei-Quinn mechanism as the explanation for why $\theta$ is small in the infrared. We study the quadratic divergences in $\theta$ arising from dimension-6 SMEFT operators and discuss the discovery prospects for these operators at electric dipole moment experiments, the LHC, and future proton-proton colliders. 
\end{abstract}

\end{center}

\vskip 1.0 cm

\end{titlepage}
\setcounter{footnote}{0} 
\setcounter{page}{1}
\setcounter{section}{0} \setcounter{subsection}{0}
\setcounter{subsubsection}{0}
\setcounter{figure}{0}



\section{Introduction}

The absence of an electric dipole moment (EDM) for the neutron strongly constrains the CP-violating QCD vacuum angle to be tiny, $\theta\lesssim 10^{-10}$~\cite{smithpurcellramsey,crewtheretal,Baker:2006ts,Afach:2015sja}. Yet CP is evidently not an exact symmetry: it is explicitly broken in the weak interactions, and must be broken further to provide the baryon asymmetry. Finding an explanation for the smallness of $\theta$ in the presence of other sources of CP violation (CPV) constitutes the strong CP problem.

Unlike the other fine-tuning problems of the Standard Model (SM), nature as we know it seems largely insensitive to $\theta$: little about nuclear physics \cite{Ubaldi:2008nf} would change if $\theta$ were of order $10^{-3}$, for example. For this reason it is widely believed that strong CP must have a dynamical, rather than anthropic, explanation, and we will make this assumption here.\footnote{For a recent attempt to connected the smallness of $\theta$ with anthropics via scanning of the cosmological constant, see~\cite{Kaloper:2017fsa}, and~\cite{Dine:2018glh} for further discussion. Also, anthropic effects could conceivably play some role in, e.g., the precise realization of the Peccei-Quinn mechanism, since axions influence the cosmological history.} 

Most of the proposed solutions to strong CP fall into two categories. In ultraviolet (UV) solutions, $\theta=0$ is a consequence of a microscopic symmetry, typically CP (Nelson-Barr models)~\cite{nelsoncp,barrcp,barrcp2,bbp} or P~\cite{Beg:1978mt,mohapatrasenjanovic,Georgi:1978xz,Babu:1989rb,barrsenjanovic}. At some intermediate scale, this symmetry is spontaneously broken, and the effects of symmetry breaking are communicated to the SM in ways that are engineered to preserve $\theta=0$. Perhaps the most compelling aspect of these models is that they take advantage of an odd property of the SM: $\theta\approx 0$ is technically natural, in the sense that radiative corrections to $\theta$ are extremely small. The first infinite and finite renormalizations induced by the CKM phase arise at 7 and 4 loop order, respectively~\cite{ellisgaillard,Khriplovich:1985jr}, and the latter has been estimated to generate $\theta\sim 4\cdot 10^{-19}$~\cite{Khriplovich:1985jr}. Thus, if $\theta=0$ can be preserved through the scale of spontaneous symmetry breaking, and the EFT at lower scales is just the SM, $\theta$ will remain sufficiently small at low scales.

In infrared (IR) solutions, $\theta$ can be absorbed into a redefinition of light fields, and the strong interactions are CP-conserving down to arbitrarily low scales. 
Under such circumstances it can also be proven that QCD does not spontaneously break CP~\cite{vafawitten}, so the strong CP problem is solved. One possibility, the massless up quark, is now strongly disfavored by lattice data~\cite{flag16}.\footnote{Although further study could still be of interest, and also provide an interesting probe of small instantons in QCD~\cite{dinedraperfestuccia}.} The remaining viable IR solution is the Peccei-Quinn (PQ) mechanism~\cite{Peccei:1977hh,Peccei:1977ur}. In this case the light field that absorbs $\theta$ is the axion $a$~\cite{Weinberg:1977ma,Wilczek:1977pj} and its potential is given by the $\theta$-dependence of the QCD vacuum energy, which is minimized at $\theta=0$ by the same theorem~\cite{vafawitten}.

Curiously, and for different reasons, neither the UV nor the IR solutions discussed above are entirely robust. A well-known example is that UV sources of explicit PQ-breaking lead to extra terms in the axion potential, stabilizing it in the wrong place and leading to a nonzero effective $\theta$~\cite{Georgi:1981pu,Lazarides:1985bj,Kamionkowski:1992mf}. To preserve $\theta\lesssim 10^{-10}$, the coefficients of Planck-scale operators must be suppressed up to high dimension ($d\sim 12$ for PQ-breaking scales of order $10^{12}$ GeV.) In other words, the PQ symmetry must be of very high quality, broken only by the QCD anomaly to great precision. This problem might be avoided with a string axion~\cite{wittenaxion}, but generally with other costs, including a nonsupersymmetric moduli problem~\cite{banksdinegraesser}. On the other hand, the most fragile component of UV solutions is the small renormalization of $\theta$ below the scale of spontaneous P/CP-breaking. If the EFT is not the SM, there can be new couplings that eventually feed the CPV spurion into $\theta$. For example, in supersymmetry, if CPV is not strongly sequestered from the SUSY-breaking sector, invariant phases in the soft parameters generate various threshold corrections to $\theta$ much larger than $10^{-10}$~\cite{dinekaganleigh,hillerschmaltz,dinedraperbn}. Likewise, in models with extra strongly-coupled gauge sectors coupled to the SM, new vacuum angles can shift $\theta$ at the confinement scale of the new sector~\cite{drapermckeen,draperkozaczukyu}. These corrections generically spoil UV solutions to strong CP.\footnote{Heavy axion solutions based on mirror $\mathbb{Z}_2$ symmetries instead of P/CP are subject to similar issues~\cite{dinedrapersuz2}. It has been argued that in cosmological supersymmetry breaking models, an $R$-axion may receive exotic contributions to its mass from interactions with the horizon that preserve the solution to strong CP~\cite{Banks:2012ty}. If this is the case, our arguments do not apply to such models.}

Experimentally, the best hope for resolving the strong CP problem is the detection of an axion component of dark matter~\cite{dinefischler,preskillwilczekwise,abbottsikivie,admx}. In contrast, UV solutions to strong CP do not make generic predictions for lower scales other than that $\theta$ should be small. However, the fragility discussed above provides another handle: UV mechanisms can still be \emph{excluded} by discovering any new physics that induces quantum corrections to $\theta$ greater than $10^{-10}$. Such a discovery would provide an upper bound on the scale at which strong CP is solved and strong indirect evidence for the PQ mechanism.

We will discuss quantum contributions to $\theta$ and associated experimental signatures in the context of the Standard Model effective theory (SMEFT). 
If integrating out heavy fields generates CPV SMEFT operators involving quarks and gluons, it will also typically produce threshold corrections to $\theta$. These corrections are not calculable in the low energy theory, but traces of them remain, including quadratically divergent corrections to $\theta$ involving the effective operators. These quadratic divergences have the same interpretation as that of the Higgs mass in the SM: they reflect strong sensitivity of the renormalizable coupling to UV physics. In the case of $\theta$ the sensitivity is cutoff-independent, $\Lambda^2/\Lambda^2$, and the relevant UV physics includes any new sources of CPV coupled to quarks or gluons. Thus, evidence for these operators sharpens the unnaturalness of small $\theta$ and strongly disfavors the possibility of natural UV solutions. We note that similar observations can be used to constrain neutrino magnetic moments based on naturalness of the neutrino masses~\cite{Voloshin:1987qy,Bell:2005kz}.

CPV SMEFT operators can be searched for at colliders and in nuclear, atomic, and molecular EDM experiments. With EDMs it is a complicated matter to extract precise values for the many Wilson coefficients involved. However, for our purposes, we need only to \emph{rule out} a bare value of $\theta$ as the only source of CPV in the strong interactions at low energies. In general this requires \emph{two} measurements, at least one of which exhibits a signal. 
Collider probes of CPV operators, on the other hand, are not ``contaminated" by $\theta$ in this way, so in principle only one measurement is required. In most cases collider sensitivity to single operators falls short of EDMs, but are still interesting, particularly for operators involving third generation quarks.

This study is organized as follows. In Sec.~\ref{sec:quad} we compute the complete quadratic divergence in $\theta$ from dimension-6 SMEFT operators. In Sec.~\ref{sec:EDM} we discuss the ability of nucleon, nuclear, and diamagnetic EDM measurements to discriminate the $\theta$-only hypothesis from $\theta$+SMEFT.  In Sec.~\ref{sec:collider} we estimate the collider sensitivity to CPV SMEFT operators, surveying the existing literature and comparing to the EDM reach. The top chromo-EDM is of particular interest, and we study the potential for the high-luminosity 14 TeV LHC and future 27 and 100 TeV colliders to detect the top cEDM in simple angular observables, where it can be distinguished from a CP-conserving magnetic moment. In Sec.~\ref{sec:summary} we summarize and conclude.

\section{Quadratic divergences and $\bar \theta$}
\label{sec:quad}

We wish to consider the combined CP violating effects of the Standard Model and some extra beyond-the-Standard-Model physics, encoded in the Lagrangian\footnote{Our conventions used in this section are collected in an appendix.}
\begin{equation}
\lag = \lag_\text{SM} + \lag_\text{BSM}.
\end{equation}
 $\lag_\text{SM}$ contains, in addition to the CKM phase, one physical strong CP phase given by the invariant combination of $\theta$ and the phases in the Yukawa couplings,
\begin{equation}
\bar \theta = \theta + \arg \det Y_u + \arg \det Y_d \, .
\end{equation}
$\lag_\text{BSM}$ may contain many new sources of CP violation. However, if the new states are sufficiently heavy, experiments will only be sensitive to a finite number of (linear combinations of) BSM phases. To leading order in momentum counting, the experimentally measurable phases are encoded in the complex Wilson coefficients of the dimension 6 SMEFT operators upon matching,\footnote{The dimension 5 Weinberg operator is not shown, as it will play no role in the following discussion.} 
\begin{equation}
\lag \stackrel{\text{low energy}}{\rightarrow} \lag_\text{SM}^\prime + \frac{1}{\Lambda^2} \sum_i c_i \op_i .
\end{equation}
Here $\Lambda$ is the mass scale of the heavy new physics, and the Wilson coefficients $c_i$ introduce up to 1149 physical phases \cite{Alonso:2013hga}, assuming conservation of baryon number. The prime of $\lag_\text{SM}^\prime$ denotes the presence of corrections to the SM parameters induced by matching and renormalization effects. In particular, the strong CP phase is shifted by
\begin{align}
\delta \bar \theta &= \delta \theta + \delta(\arg\det Y_u) + \delta(\arg\det Y_d) \\
&\approx \delta \theta + \mathrm{Im Tr} (Y_u^{-1} \delta Y_u) + \mathrm{Im Tr} (Y_d^{-1} \delta Y_d) 
\label{eq:deltathetabar}
\end{align}
where in the second line we have expanded to first order in the threshold corrections $\delta Y_u$ and $\delta Y_d$.

Absent specific knowledge of the form of $\lag_\text{BSM}$, $\{\delta \theta, \delta Y_u, \delta Y_d\}$ are incalculable. However, one can estimate their natural size from SMEFT loops. Just as quadratic divergences in the Higgs mass from loops of SM fields signals strong sensitivity of $m_H^2$ to UV threshold corrections, quadratically divergent corrections to $\theta$ and the quark Yukawas in SMEFT are a proxy for the threshold corrections received by these parameters at the cutoff. 
The one loop quadratically-sensitive corrections to the SM parameters are
\begin{align}
\delta \theta &\sim \frac{1}{\Lambda^2} \Big(  \frac{2}{g_s^2} c_{H\tilde G}  - \frac{9}{2 g_s} c_{\tilde G} \Big) \Lambda^2  \\
(\delta Y_d)^{ij} &\sim \frac{1}{16 \pi^2 \Lambda^2} \Big(  3 c_{dH}^{ij} - ( c_{Hq(1)}^{ik} + 3 c_{Hq(3)}^{ik} ) Y_d^{kj} + Y_d^{ik} c_{Hd}^{kj} - Y_u^{ik} c_{Hud}^{kj} & \nonumber \\
&~~~~~~~~~~~~~~~~+ 4(c_{qd(1)}^{jmni} + \frac{4}{3} c_{qd(8)}^{jmni} ) Y_d^{mn} - 2 c_{ledq}^{*mnji} Y_e^{mn} \nonumber \\
&~~~~~~~~~~~~~~~~ + (6 c_{quqd(1)}^{mnij} + c_{quqd(1)}^{inmj} + \frac{4}{3} c_{quqd(8)}^{inmj}) Y_u^{\dagger nm} \nonumber \\
&~~~~~~~~~~~~~~~~ + g^\prime c_{dB}^{ij} - 18 g c_{dW}^{ij} - 16 g_s c_{dG}^{ij} \Big) \Lambda^2 &\\ 
(\delta Y_u)^{ij} &\sim \frac{1}{16 \pi^2 \Lambda^2} \Big(  3 c_{uH}^{ij} + ( c_{Hq(1)}^{ik} - 3 c_{Hq(3)}^{ik} ) Y_u^{kj} - Y_u^{ik} c_{Hu}^{kj} + Y_d^{ik} c_{Hud}^{* jk} &\nonumber \\
&~~~~~~~~~~~~~~~~+ 4(c_{qu(1)}^{jmni} + \frac{4}{3} c_{qu(8)}^{jmni} ) Y_u^{mn} + 2 c_{lequ(1)}^{mnij} Y_e^{\dagger nm}  \nonumber \\
&~~~~~~~~~~~~~~~~ + (6 c_{quqd(1)}^{ijmn} + c_{quqd(1)}^{mjin} + \frac{4}{3} c_{quqd(8)}^{mjin}) Y_d^{\dagger nm}  \nonumber \\
&~~~~~~~~~~~~~~~~ - 5 g^\prime c_{uB}^{ij} - 18 g c_{uW}^{ij} - 16 g_s c_{uG}^{ij} \Big) \Lambda^2  .
\end{align}
The $c$s are the dimensionless Wilson coefficients of dimension 6 operators, including a variety of electromagnetic, weak, and chromo EDMs, four-fermi operators, the Weinberg operator, and various $d=4$ operators with $H^\dagger H$ attached. 
The overall correction to $\bar \theta$ is 
\begin{align}
16 \pi^2 \delta \bar \theta \sim & 16 \pi^2 \Big(  \frac{2}{g_s^2} c_{H\tilde G}  - \frac{9}{2 g_s} c_{\tilde G} \Big) \nonumber \\
& + \Im\, \Tr [Y_d^{-1} (3 c_{dH}  + g^\prime c_{dB} - 18 g c_{dW} - 16 g_s c_{dG} ) ] \nonumber \\
& + \Im\, \Tr [Y_u^{-1} (3 c_{uH}  -5 g^\prime c_{uB} - 18 g c_{uW} - 16 g_s c_{uG} ) ] \nonumber \\
& + \Im\, \Tr [(Y_d^{-1} Y_u + Y_d^\dagger (Y_u^\dagger)^{-1}) c_{Hud}] \nonumber \\ 
& + \Im [ 2 c_{lequ(1)}^{mnij} Y_e^{\dagger nm} (Y_u^{-1})^{ji} - 2 c_{ledq}^{*mnij} Y_e^{mn}(Y_d^{-1})^{ij}] \nonumber \\
& + \Im [(6 c_{quqd(1)}^{mnij} + c_{quqd(1)}^{inmj} + \frac{4}{3} c_{quqd(8)}^{inmj}) (Y_u^{\dagger nm}(Y_d^{-1})^{ji} + Y_d^{\dagger ji}(Y_u^{-1})^{nm} )].
\label{eq:thetacorr}
\end{align}
The contributions of $c_{Hq(1)},c_{Hq(3)},c_{Hu},c_{Hd},c_{qu(1)},c_{qu(8)},c_{qd(1)},c_{qd(8)}$, which appear in $\delta Y_{u,d}$, vanish identically in $\delta\bar\theta$ due to Hermiticity. The operators whose coefficients appear explicitly in Eq.~(\ref{eq:thetacorr}) are listed  in  Table \ref{fig:optable}, and a sampling of the diagrams are shown in Fig.~\ref{fig:thetadiag}. All the loop integrals were regulated with a momentum space cutoff $\Lambda$. In the case of the Weinberg ($\op_{\tilde G}$) and dipole operators ($\op_{[u/d][B/W/G]}$), the loop integrals are scaleless; for the remaining operators, we have checked our results against the dimensionally regularized RG equations of \cite{Jenkins:2013zja}.

\begin{table}[t!]
\begin{center}
\begin{tabular}{c | c || c | c}
$\op_{uH}$ & $H^\dagger H \overline{Q_{Li}} \tilde H u_{Rj}$ &
$\op_{dH}$ & $H^\dagger H \overline{Q_{Li}} H d_{Rj}$ \\
$\op_{dG}$ & $\overline{Q_{Li}} \sigma^{\mu\nu} T^a d_{Rj} H G^a_{\mu\nu}$ &
$\op_{dW}$ & $\overline{Q_{Li}} \sigma^{\mu\nu} d_{Rj} \tau^a H W^a_{\mu\nu}$ \\
$\op_{dB}$ & $\overline{Q_{Li}} \sigma^{\mu\nu} d_{Rj} H B_{\mu\nu}$ &
$\op_{uG}$ & $\overline{Q_{Li}} \sigma^{\mu\nu} T^a u_{Rj} \tilde H G^a_{\mu\nu}$ \\
$\op_{uW}$ & $\overline{Q_{Li}} \sigma^{\mu\nu} u_{Rj} \tau^a \tilde H W^a_{\mu\nu}$ &
$\op_{uB}$ & $\overline{Q_{Li}} \sigma^{\mu\nu} u_{Rj} \tilde H B_{\mu\nu}$ \\
$\op_{Hud}$ & $i \tilde H^\dagger D_\mu H \overline{u_{Ri}} \gamma^\mu d_{Rj}$ &
$\op_{quqd(1)}$ & $\epsilon^{ef} \overline{Q^e_{Li}} u_{Rj} \overline{Q^f_{Lk}} d_{Rl}$ \\
$\op_{quqd(8)}$ & $\epsilon^{ef} \overline{Q^e_{Li}} T^a u_{Rj} \overline{Q^f_{Lk}} T^a d_{Rl}$ &
$\op_{lequ(1)}$ & $\epsilon^{ef} \overline{L^e_{Li}} e_{Rj} \overline{Q^f_{Lk}} u_{Rl}$ \\
$\op_{ledq}$ & $\overline{L_{Li}} e_{Rj} \overline{d_{Rk}} Q_{Ll}$ &
$\op_{H \tilde G}$ & $H^\dagger H G^a_{\mu\nu} \widetilde G^{a \mu\nu}$ \\
$\op_{\tilde G}$ & $f^{abc} G^{a\mu}_{~~\nu} G^{b\nu}_{~~\rho} \widetilde G^{c\rho}_{~~\mu}$
\end{tabular}
\end{center}
\caption{The dimension 6 operators of the Standard Model (in the basis of Ref.~\cite{Grzadkowski:2010es}) which contribute to the one loop quadratic divergence in $\bar \theta$. 
$\epsilon^{12} = \epsilon_{12} = +1$ and $\sigma^{\mu\nu} = \frac{1}{2} i [\gamma^\mu,\gamma^\nu]$.\label{fig:optable}}
\end{table}
Absent an infrared relaxation of $\bar\theta$, e.g.~by the Peccei-Quinn mechanism, 
naturalness requires $\abs{\delta \bar \theta} \lesssim 10^{-10}$, implying a stringent bound on the combination of Wilson coefficients in (\ref{eq:thetacorr}).

We see that $\delta \bar \theta$ receives contributions from a considerable variety of operators, all containing colored particles but many also containing leptons and electroweak bosons. If one assumes the Wilson coefficients are minimally flavor-violating \cite{DAmbrosio:2002vsn}, i.e.~at leading order in the Yukawas
\begin{align}
c_{d[H/B/W/G]}^{ij} &= \hat c_{d[H/B/W/G]} Y_d^{ij}; \quad
c_{u[H/B/W/G]}^{ij} = \hat c_{u[H/B/W/G]} Y_u^{ij}; \quad
c_{Hud}^{ij} = \hat c_{Hud} (Y_u^\dagger Y_d)^{ij}; \nonumber \\
c_{lequ(1)}^{ijkl} &= \hat c_{lequ(1)} Y_e^{ij} Y_u^{kl}; \quad
c_{ledq}^{ijkl} = \hat c_{ledq} Y_e^{ij} (Y_d^\dagger)^{kl}; \nonumber \\
c_{quqd[(1)/(8)]}^{ijkl} &= \hat c_{quqd[(1)/(8)]}^{A} Y_u^{ij} Y_d^{kl} + \hat c_{quqd[(1)/(8)]}^{B} Y_u^{kj} Y_d^{il},
\end{align}
then Eq.~(\ref{eq:thetacorr}) reduces to a sum over the hatted flavor-blind phases
\begin{align}
16 \pi^2 \delta \bar \theta |_\text{MFV} \sim & 16 \pi^2 \Big( 2 c_{H\tilde G}  + 9 g_s c_{\tilde G} \Big) \nonumber \\
& + 3 \Im [3 \hat c_{dH}  + g^\prime \hat c_{dB} - 18 g \hat c_{dW} - 16 g_s \hat c_{dG} + 3 \hat c_{uH}  -5 g^\prime \hat c_{uB} - 18 g \hat c_{uW} - 16 g_s \hat c_{uG}] \nonumber \\
& + (\Tr[Y_u Y_u^\dagger]+\Tr[Y_d Y_d^\dagger]) \Im [ 7 \hat c_{quqd(1)}^{A} + 7 \hat c_{quqd(1)}^{B} + 7 \hat c_{quqd(8)}^{A} + 7 \hat c_{quqd(8)}^{B} - \hat c_{Hud}] \nonumber \\ 
& + \Tr[Y_e Y_e^\dagger] \Im [ 2 \hat c_{ledq} + 2 \hat c_{lequ(1)}]. 
\end{align}
However, if the new physics has a different flavor structure, it need not even be CP violating to give a sizable contribution to $\delta \bar \theta$: the presence of the CKM phase in (\ref{eq:thetacorr}) will often suffice.

Why might this bound on the Wilson coefficients be satisfied? One, the BSM physics may couple extremely weakly to the SM, either through small couplings or through suppression by a large number of loop factors. For example, requiring that the Wilson coefficients be suppressed by a factor of
\begin{equation}
\left( \frac{1}{16 \pi^2} \right)^n \sim \left( \frac{g^{\prime2}}{16 \pi^2} \right)^{n^\prime} \sim 10^{-10}
\end{equation}
implies loop orders of $n \sim 4.5$ and $n^\prime \sim 3.1$. However, in this case, other effects from BSM physics would also be too small to observe.

Another possibility is that the Wilson coefficients may be of natural size, but their combination in Eq.~(\ref{eq:thetacorr}) is very small (analogous to a Veltman condition for the Higgs mass.) This apparent fine tuning would still require explanation, especially as it is not stable under the 1-loop SMEFT RGEs (Eq.~(\ref{eq:thetacorr}) is not an eigendirection of the anomalous dimension matrix of the dimension 6 operators)~\cite{Alonso:2013hga}. This RGE instability also highlights another problem: there are two-loop and higher corrections to $\bar \theta$ from other dimension 6, and higher, operators which we have neglected, and may yet be significant as well.\footnote{For one example, the $\op_{lequ}^{(3)}$ operator does not appear in Table~\ref{fig:optable}, but it generates $\op_{lequ}^{(1)}$ under RG, which does appear in the table. We will comment further on $\op_{lequ}^{(1,3)}$ in Sec.~\ref{sec:EDM}.}

Thus, even if BSM physics is too heavy to be produced on-shell, observation of a nonzero SMEFT coefficient in Eq.~(\ref{eq:thetacorr}) would cause significant tension for UV solutions to the strong CP problem. In the following sections, we discuss sensitivities of low-energy and collider experiments to some representative operators.

\tikzset{scalar/.style={dashed}}
\tikzset{fermion/.style={
        decoration={markings,
            mark= at position 0.65 with {\arrow[scale=2.5]{latex}} ,
        },
        postaction={decorate}
    }
}
\tikzset{dim6blob/.style={circle,draw=black,thick,minimum size=25pt,fill=black,pattern= north west lines}}
\tikzset{vector/.style={decorate, decoration={snake,segment length=5pt,amplitude=3pt}}}
\begin{figure}
\centering
\begin{tikzpicture}[scale=1.6]
\coordinate (q) at (-1,0);
\coordinate (u) at (1,0);
\node[dim6blob,label=south east:{$\op_{uH},\op_{dH}$}] (v) at (0,0) {};
\coordinate (h) at (0,-0.8);
\draw[fermion] (q) -- (v);
\draw[fermion] (v) -- (u);
\draw[scalar] (v) -- (h);
\draw[scalar] (v) to [in=140,out=40,looseness=7] (v);
\end{tikzpicture} 
\begin{tikzpicture}[scale=1.6]
\coordinate (q) at (-1,0);
\coordinate (halfq) at (-0.65,0);
\coordinate (u) at (1,0);
\node[dim6blob,label=south east:{$\op_{Hud}$}] (v) at (0,0) {};
\coordinate (h) at (0,-0.8);
\draw (q) -- (halfq);
\draw[fermion] (halfq) -- (v);
\draw[fermion] (v) -- (u);
\draw[scalar] (v) -- (h);
\draw[scalar] (v) to [in=-90,out=-135,looseness=2] (halfq);
\end{tikzpicture}  
\begin{tikzpicture}[scale=1.6]
\coordinate (q) at (-1,0);
\coordinate (halfq) at (-0.65,0);
\coordinate (u) at (1,0);
\node[dim6blob,label=south east:{\pbox[t]{10mm}{$\op_{u[B/W/G]}$, \\ $\op_{d[B/W/G]}$}}] (v) at (0,0) {};
\coordinate (h) at (0,-0.8);
\draw (q) -- (halfq);
\draw[fermion] (halfq) -- (v);
\draw[fermion] (v) -- (u);
\draw[scalar] (v) -- (h);
\draw[vector] (v) to [in=-90,out=-135,looseness=2] (halfq);
\end{tikzpicture} 
\begin{tikzpicture}[scale=1.6]
\coordinate (q) at (-1,0);
\coordinate (halfq) at (0,-0.6);
\coordinate (u) at (1,0);
\node[dim6blob,label={[label distance=10pt]-15:{\pbox[t]{10mm}{$\op_{ledq},\op_{lequ(1)}$, \\ $\op_{quqd[(1)/(8)]}$}}}] (v) at (0,0) {};
\coordinate (h) at (0,-0.8);
\draw[fermion] (q) -- (v);
\draw[fermion] (v) -- (u);
\draw[scalar] (halfq) -- (h);
\draw (v) to [in=180,out=-145,looseness=2] (halfq);
\draw (halfq) to [in=-35,out=0,looseness=2] (v);
\end{tikzpicture} \\
\begin{tikzpicture}[scale=1.6]
\coordinate (gm) at (-0.8,0);
\coordinate (halfm) at (-0.4,0);
\coordinate (gp) at (0.8,0);
\node[dim6blob,label={south east:{$\op_{\tilde G}$}}] (halfp) at (0.4,0) {};
\draw[vector] (gm) -- (halfm);
\draw[vector] (halfp) -- (gp);
\draw[vector] (halfp) to [in=90,out=135,looseness=2] (halfm);
\draw[vector] (halfp) to [in=-90,out=-135,looseness=2] (halfm);
\end{tikzpicture}
\begin{tikzpicture}[scale=1.6]
\coordinate (gm) at (-0.8,0);
\coordinate (gp) at (0.8,0);
\node[dim6blob,label={south east:{$\op_{H \tilde G}$}}] (v) at (0,0) {};
\draw[vector] (gm) -- (v);
\draw[vector] (v) -- (gp);
\draw[scalar] (v) to [in=140,out=40,looseness=7] (v);
\end{tikzpicture} 
\caption{Diagrammatic examples of the one-loop contributions of the dimension 6 operators (indicated by the hatched circle) to the dimension 4 strong CP phase.\label{fig:thetadiag}}
\end{figure}
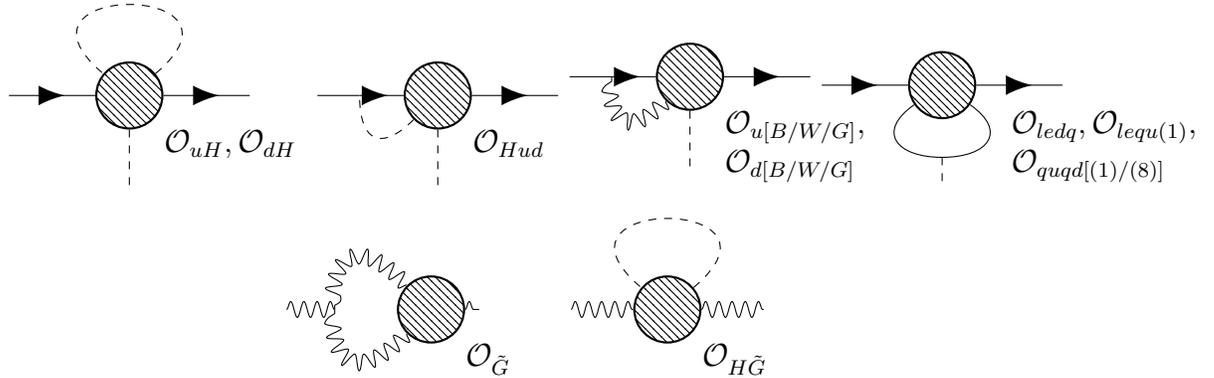

\section{Low-energy probes}
\label{sec:EDM}
The main observational consequence of a nonzero $\bar \theta$ term is the presence of nonzero EDMs of nucleons, nuclei, atoms, and molecules. The absence of a signal in all neutron-EDM experiments thus far provides the limit $\bar \theta \lesssim 10^{-10}$ and establishes the strong CP problem. A signal in any of the next-generation EDM experiments, however, might point toward a finite $\bar \theta$ term, higher-dimensional BSM operators, or a combination of both.\footnote{The CKM phase contributes to EDMs at a level significantly below current and expected future experimental sensitivities, and can be neglected.} Strategies to isolate the source of CP violation from multiple nonzero EDM measurements have appeared in the literature \cite{Lebedev:2004va,jordyunraveling}. In particular, a scenario with a pure $\bar \theta$ term would lead to a rather distinct pattern of nucleon, nuclear, and diamagnetic EDMs (from now on, we will refer to these as `hadronic' EDMs) \cite{deVries:2011an, Bsaisou:2012rg}, while lepton and paramagnetic EDMs, that are dominated by (semi-)leptonic sources of CP violation, should be much smaller \cite{Choi:1990cn, Ghosh:2017uqq}. While it is challenging to completely isolate the source of a new observation of hadronic CP violation, it is potentially  easier to rule out a pure $\bar\theta$ scenario. As discussed above, such a result can provide indirect evidence that $\bar\theta$ is relaxed by the PQ mechanism.\footnote{Unfortunately, discovering  EDMs consistent with $\bar\theta$ would provide less information. For example, it would not rule out the PQ mechanism, since, as discussed in the introduction, some level of explicit PQ-violation is expected from UV sources. Furthermore, in the presence of higher-dimensional BSM sources of CP violation, such as quark chromo-EDMs \cite{Pospelov:2000bw} or certain CP-odd four-quark operators \cite{Cirigliano:2016yhc}, the PQ mechanism does not relax $\bar \theta$ to zero, but instead to a finite value proportional to the Wilson coefficients of the BSM operators. Depending on the details of the setup, the induced $\bar \theta$ term can potentially dominate hadronic EDMs. See e.g. Ref.~\cite{Maiezza:2014ala} for an explicit realization in a left-right symmetric model.}

\subsection{$\bar\theta$ dependence of EDMs}
In testing the consistency of putative EDM signals with the SM + nonzero $\bar\theta$, there are theoretical challenges. Non-perturbative QCD and nuclear- and atomic-structure calculations are required to link  $\bar \theta$  to EDMs  of nucleons, nuclei, and diamagnetic atoms. Nevertheless, in recent years much progress has been made which we  summarize here. 

Historically the most important EDM is that of the neutron. $d_n$ has been the target of many experiments for over six decades, leading to the present limit $|d_n| < 3.0\cdot 10^{-13}$ $e$ fm \cite{Baker:2006ts,Afach:2015sja}. The first genuine calculation of the neutron EDM in terms of $\bar \theta$ was performed in Ref.~\cite{crewtheretal} using current algebra techniques and coincides with a leading-order calculation in chiral perturbation theory ($\chi$PT)\cite{Hockings:2005cn}. At next-to-leading order in $\chi$PT the neutron EDM is given by \cite{Mereghetti:2010kp}
\begin{equation}\label{neutronEDM}
  d_n = {\bar d}_n - \frac{e g_A\bar g_0}{8\pi^2 F_\pi} \left(  \ln
\frac{m_\pi^2}{m_N^2} -\frac{\pi m_\pi}{2 m_N} \right)\,,
\end{equation}
in terms of $g_A \simeq 1.27$, the strong pion-nucleon axial coupling, $F_\pi \simeq 92.2$ MeV, the pion decay constant, $m_\pi$, the pion mass, and two low-energy constants (LECs), $\bar d_n$ and $\bar g_0$, associated to CP-violating hadronic interactions that are introduced below. The expression for the proton EDM is, not surprisingly, very similar,
\begin{equation}\label{protonEDM}
d_p = {\bar d}_p +\frac{e g_A}{8\pi^2 F_\pi} \left[ \bar g_0 \left(  \ln
\frac{m_\pi^2}{m_N^2} -\frac{2\pi m_\pi}{m_N} \right) -
\bar g_1 \frac{\pi m_\pi}{2 m_N} \right]\,,
\end{equation}
and depends on two additional LECs  $\bar d_p$ and $\bar g_1$.

The LECs $\bar g_0$ and $\bar g_1$ are the coupling constants of CP-violating pion-nucleon interactions
\begin{equation}
\mathcal L_{\pi N} = \bar g_0\,\bar N \vec \tau \cdot \vec \pi N\, + \bar g_1\,\bar N \pi_3 N\,,
\end{equation} 
in terms of the nucleon doublet $N = (p\,n)^T$ and pion triplet $\vec \pi$. The logarithm in brackets in Eqs.~\eqref{neutronEDM} and \eqref{protonEDM} arise from one-loop diagrams involving one insertion of $\bar g_0$, one insertion of the strong pion-nucleon coupling $g_A$, and a photon coupling to the pion-in-flight. This loop is divergent and the divergence and associated scale dependence is absorbed into the counterterms $\bar d_n$ and $\bar d_p$, which reflect short-distance (of distance shorter than $\sim m_\pi^{-1}$) contributions to the nucleon EDMs. The other pieces in brackets arise from finite loops at next-to-leading order in the chiral expansion.

The above CP-odd hadronic interactions $d_n$, $d_p$, $\bar g_0$, and $\bar g_1$ also determine the EDMs of light nuclei and diamagnetic atoms in the pure $\bar \theta$ scenario.  Other interactions, such as short-range CP-odd nucleon-nucleon couplings, only appear at next-to-next-to-leading order in the chiral expansion and are expected to contribute at the $10\%$ level \cite{deVries:2012ab, Bsaisou:2014oka}. So far there exist no EDM experiments involving light nuclei or light atoms. Light nuclei have not been used because they are charged, and standard EDM experiments apply a large electric field which would eject the nucleus from the apparatus. Light atoms are not used because Schiff's theorem \cite{Schiff:1963zz} ensures that the EDM of a nucleus is screened inside a neutral system such as an atom. Schiff's theorem is not exact and assumes point-like particles, but provides a very good approximation for small systems. As such, light atoms are not appropriate targets for EDM searches and much heavier systems are used. 

The current best limit on any EDM is that of the ${}^{199}$Hg atom: $d_{\mathrm{Hg}} < 6.2 \cdot 10^{-17}$ $e$ fm \cite{Graner:2016ses}. Unfortunately, it is not an easy task to calculate the atomic EDM of such a complex system in terms of the above CP-odd interactions. Technically, it requires a calculation of the so-called nuclear Schiff moment and an atomic calculation linking the Schiff moment to the atomic EDM. At present, the atomic calculation is under relatively good control \cite{Yamanaka:2017mef, Sahoo:2018ile}, but the nuclear calculation is problematic (see Refs.~\cite{Engel:2013lsa, Yamanaka:2017mef} for more details). The present estimate is
\begin{eqnarray}\label{dHg}
d_{\rm Hg}&=& -(1.8\pm 0.3)\cdot 10^{-4}\bigg[(1.9\pm0.1)d_n +(0.20\pm 0.06)d_p\nonumber\\
&&\hspace{3cm}+\bigg(0.13^{+0.5}_{-0.07}\,\bar g_0 + 0.25^{+0.89}_{-0.63}\,\bar g_1\bigg)e\, {\rm fm}\bigg]\,,
\end{eqnarray}
where the term in front of the brackets is the atomic Schiff screening factor. The main problem is the size (and even sign, in the case of $\bar g_1$) of the coefficients in front of the CP-odd pion-nucleon couplings, which are very uncertain. Advances in nuclear theory are required to improve these calculations. In specific BSM scenarios there can be other sizable contributions to $d_{\mathrm{Hg}}$, for example from (semi-)leptonic CP-odd interactions, but these are negligible in the pure $\bar \theta$ scenario. The EDM of the ${}^{129}$Xe atom can be considered along similar lines but suffers from a larger screening factor and similar nuclear uncertainties, while the experimental limit is not as stringent \cite{PhysRevLett.86.22}.  Therefore we do not consider it here.

An interesting system is the ${}^{225}$Ra atom, the EDM of which has been bounded by $d_{\text{Ra}} < 1.2 \cdot 10^{-10}$ $e$ fm \cite{Bishof:2016uqx}. While this limit is seven orders of magnitude weaker than that on $d_{\mathrm{Hg}}$, great experimental progress is expected. In addition, the atomic screening factor is less severe for this atom and, more importantly, due to its octopole-deformed shape, the coefficients in front of $\bar g_0$ and $\bar g_1$ are significantly enhanced with respect to Hg:
\begin{equation}\label{dRa}
d_{\mathrm{Ra}} = (7.7\pm 0.8)\cdot 10^{-4}\cdot\left[(2.5\pm 7.5)\,\bar g_0 - (65 \pm 40)\,\bar g_1\right]e\, {\rm fm}\,.
\end{equation}
While the nuclear uncertainties are still significant, they are under relatively better control than for Hg \cite{Engel:2013lsa,Dobaczewski:2018nim}.

As discussed above, EDM experiments traditionally involve neutral systems. However, it was realized that charged particles trapped in electromagnetic storage rings can also be used \cite{Farley:2003wt}. In this way, the g-2 collaboration set the first limit on the muon EDM \cite{Bennett:2008dy}. Several experimental collaborations aim to construct storage rings to measure the EDMs of the proton and deuteron and perhaps even the ${}^3$He nucleus. Great progress towards these measurements have been reported in Refs.~\cite{Eversmann:2015jnk, Guidoboni:2016bdn} and it has been claimed that an accuracy of $10^{-16}$ $e$ fm can be achieved in such a setup. While still less precise than the $d_{\text{Hg}}$ measurement it must be stressed that light nuclei would not suffer from atomic screening nor from large nuclear uncertainties. These plans have lead to  considerable activity in the nuclear community and the EDMs of several light nuclei have been calculated within the framework of chiral effective field theory \cite{deVries:2011an, Bsaisou:2014zwa}
\begin{eqnarray} \label{eq:h2edm} 
d_{{}^2\text{H}} &=&
(0.94\pm0.01)(d_n + d_p) + \bigl [ (0.18 \pm 0.02) \,\bar g_1\bigr] \,e \,{\rm fm} \, ,\\
d_{{}^3\text{He}} &=& (0.90\pm0.01)d_n -(0.03 \pm0.01) d_p \nonumber\\
				&&+ \left[ (0.11\pm0.01) \bar g_0 + (0.14\pm0.02) \bar g_1\right] \,e  \,{\rm fm} \, .
\end{eqnarray}
EDMs of other light nuclei such as ${}^6$Li, ${}^9$Be, and ${}^{13}$C have been calculated in terms of the same LECs using a nuclear cluster model \cite{Yamanaka:2015qfa,Yamanaka:2016umw}. The results indicate that such systems do not show large enhancements or suppression with respect to ${}^2$H and ${}^3$He EDMs.\footnote{For brevity, in what follows we refer to the ${}^2$H and ${}^3$He EDMs as $d_D$ and $d_{\rm He}$, respectively.}

The above relations show that we can calculate a handful of EDMs of experimental interest in terms of four hadronic CP-violating coupling constants. The missing link is the calculation of $d_n$, $d_p$, $\bar g_0$, and $\bar g_1$ in terms of $\bar \theta$. By far, the size of $\bar g_0$ is known to the highest accuracy. The $\bar \theta$ term can, via the axial $U(1)$ anomaly, be rotated into a complex quark mass. As such, hadronic interactions induced by $\bar \theta$ are linked to hadronic interactions induced by the CP-conserving quark mass terms \cite{Mereghetti:2010tp}. This was already appreciated in Ref.~\cite{crewtheretal} and $\bar g_0$ was linked to a linear combination of octet baryon masses. Recently it was realized that this relation is badly violated at higher orders in the chiral $SU(3)$ expansion and that the only reliable relation is between $\bar g_0$ and the strong proton-neutron mass splitting \cite{deVries:2015una}. As the strong proton-neutron mass splitting has been a target of various lattice calculations it is known to high accuracy \cite{Brantley:2016our} and we obtain
\begin{equation}\label{g0theta}
\bar g_0 = -(14.7\pm2.3)\cdot 10^{-3}\,\bar \theta\,.
\end{equation}

Unfortunately a relation with comparable precision does not exist for $\bar g_1$. The main difficulty is that $\bar g_1$ is an isospin-breaking interaction while the $\bar \theta$ term conserves isospin. As such, $\bar g_1$ is not directly induced by $\bar \theta$ but only via interplay with isospin breaking via the quark masses. This obscures the link between $\bar g_1$ and the hadron mass spectrum which is so useful in case of $\bar g_0$. Nevertheless, a piece of $\bar g_1$ can be linked to the strong pion mass splitting. The remaining piece is unknown but has been estimated in a model in Ref.~\cite{Bsaisou:2012rg} where it was found to be relatively small. Adding this piece as an additional uncertainty, we obtain \cite{deVries:2015una}
\begin{equation}\label{g1theta}
\bar g_1 = (3.4\pm2.4)\cdot 10^{-3}\,\bar \theta\,.
\end{equation}
The smallness of $|\bar g_1/\bar g_0|$ can be understood from the necessity of additional isospin breaking for $\bar g_1$. 

Finally, we need to know the values of the nucleon EDMs. An estimate can be given by inserting the obtained values of $\bar g_0$ and $\bar g_1$ in Eqs.~\eqref{neutronEDM} and \eqref{protonEDM}. This gives
\begin{equation}
d_n = \bar d_n - (2.1\pm0.3)\cdot 10^{-3}\,\bar \theta\,e\,\mathrm{fm},\qquad d_p = \bar d_p + (2.4\pm0.3)\cdot 10^{-3}\,\bar \theta\,e\,\mathrm{fm}\, ,
\end{equation}
which can be used as an estimate if it is assumed, which is often done, that the short-distance contributions $\bar d_n$ and $\bar d_p$ are small with respect to the chiral logarithm. However, chiral techniques do not allow for a solid estimate of the nucleon EDMs due to the unknown sizes of $\bar d_n$ and $\bar d_p$. Non-perturbative techniques are required. Refs.~\cite{Pospelov:1999ha,Hisano:2012sc} calculated the neutron EDM directly using QCD sum rules and found
\begin{equation}
d_n(\text{QCD sum rules}) = -(2.4 \pm 1.2)\cdot 10^{-3}\,\bar \theta\,e\,\mathrm{fm}\,,
\end{equation}
in reasonable agreement with the chiral estimate. 

The proton EDM is expected to be of the same magnitude as the neutron EDM but with opposite sign. The sum rules analyzed in~\cite{Pospelov:1999ha,Hisano:2012sc} suggest $d_p\approx -3/2 d_n$, while a recent calculation using a large $N_c$ QCD model and gauge/string duality found
$d_n = - d_p = -1.8 \cdot 10^{-3}\,\bar \theta\,e\,\mathrm{fm}$ without an uncertainty estimate \cite{Bartolini:2016jxq}.

Ideally, the nucleon EDMs would be calculated with lattice QCD techniques, and in recent years several collaborations have attempted to do so \cite{Guo:2015tla, Shindler:2015aqa, Alexandrou:2015spa, Shintani:2015vsx}. Very accurate results at non-physical pion masses were, for example, reported in Refs. \cite{Guo:2015tla} and \cite{Alexandrou:2015spa} and an extrapolation to the physical point of the data in Ref.~\cite{Guo:2015tla} lead to $d_n = -(3.9 \pm 0.9)\bar \theta\,e\,\mathrm{fm}$. Unfortunately, it was recently argued that all existing lattice calculations suffered from spurious EDM contributions due to mixing with the CP-even anomalous magnetic moment \cite{Abramczyk:2017oxr}. Subtracting the spurious pieces lead to lattice signals consistent with zero with uncertainties larger than the model estimates given above. This implies that current lattice calculations are not yet precise enough  to accurately calculate the nucleon EDMs with a small nonzero $\bar \theta$. Further work is required; see, for example, Refs.~\cite{Abramczyk:2018dqb, Dragos:2018uzd}. 

In our discussion below, we will use the QCD sum rules calculation of $d_n$ and set $d_p = -(1\pm 0.5) d_n$, which essentially covers all existing estimates. However, our numerical results can also simply be regarded as an illustration, and can be straightforwardly updated without modifying the qualitative point if more precise calculations become available in the future. We express the EDMs of light nuclei and diamagnetic atoms in terms of $d_n$, $d_p$ and $\bar g_{0,1}$ via the relations given above and use Eqs.~\eqref{g0theta} and \eqref{g1theta} to link $\bar g_{0,1}$ to $\bar \theta$. 

\subsection{Excluding pure-$\bar\theta$ with correlated measurements}
\begin{figure}
\begin{center}
\includegraphics[scale = .55]{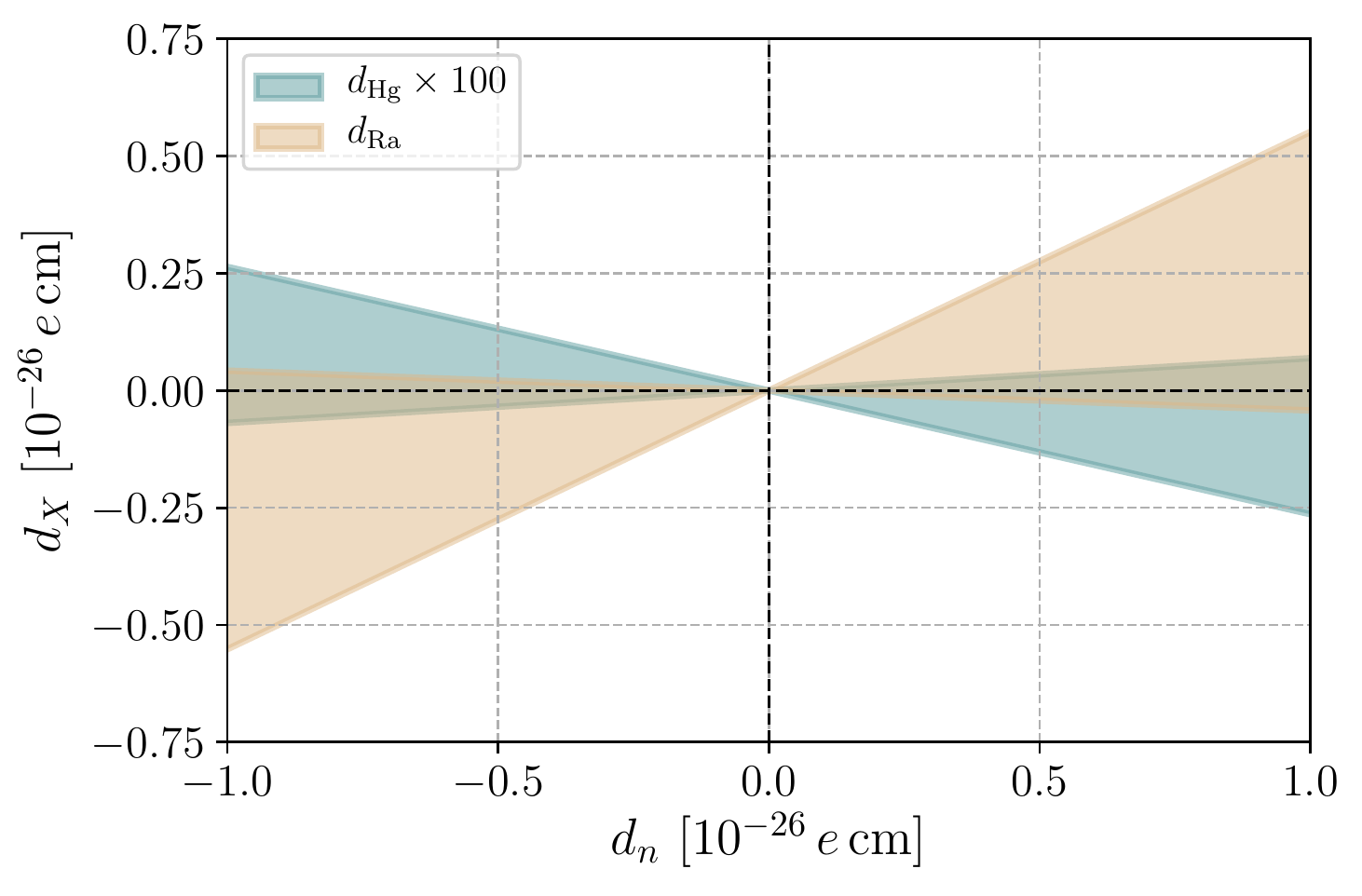}
\includegraphics[scale = .55]{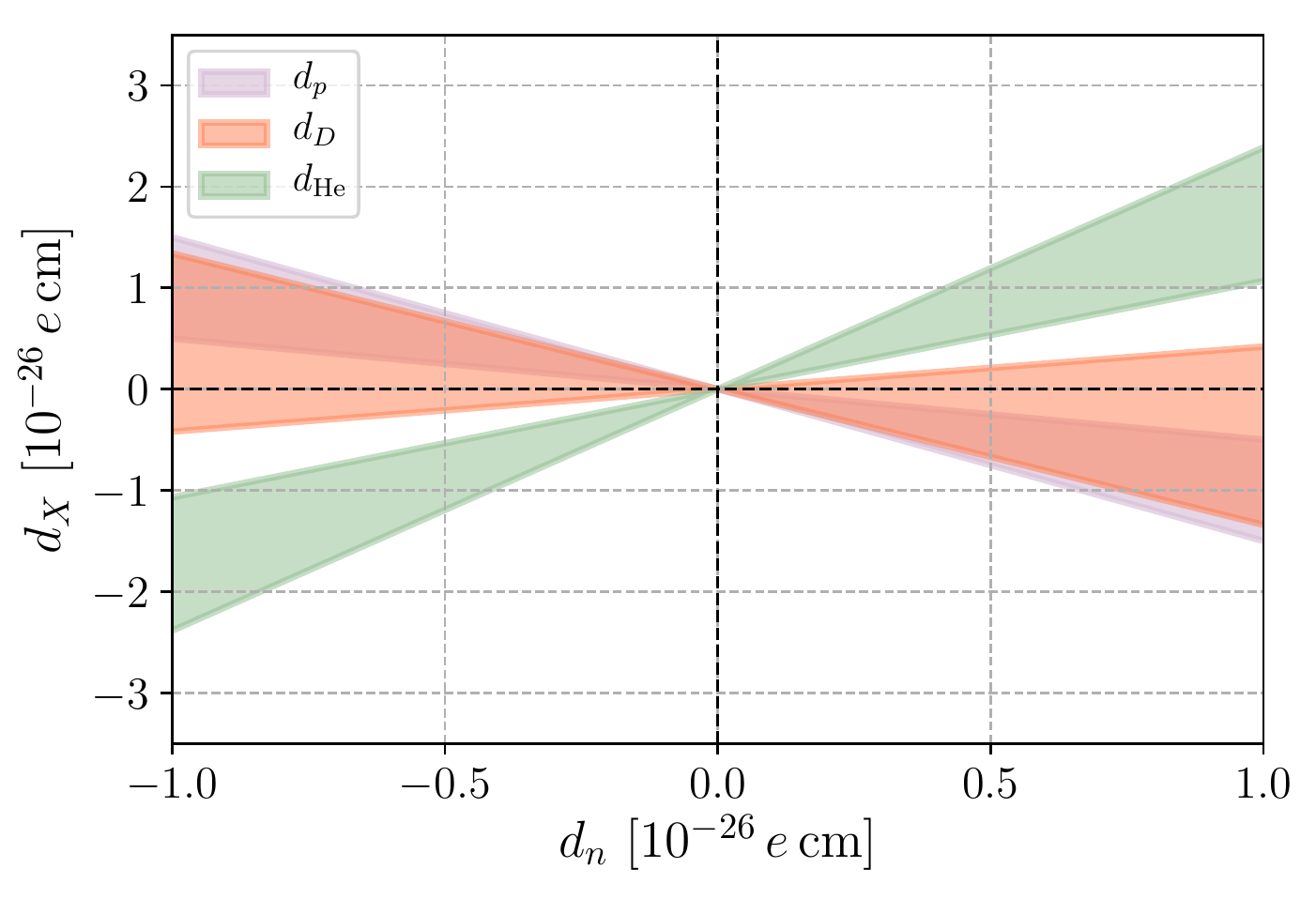}
\end{center}
\caption{Values of various EDMs as function of the neutron EDM that are consistent with a pure $\bar \theta$ scenario. Any EDM measurement outside of any of the shaded regions would point towards BSM sources of CP violation, indicating that the strong CP problem very likely requires an infrared solution.}
\label{EDMcorrelation}
\end{figure}

To rule out a pure $\bar \theta$ scenario, we need either a single measurement of an EDM of a leptonic or paramagnetic system which would hint at a (semi-)leptonic source of CP violation, or at least two hadronic EDM measurements whose relative size is in conflict with the relations above. While (semi-)leptonic CPV would rule out a pure $\bar \theta$ scenario, it would not immediately point towards a PQ mechanism, since some dimension-six  CP-violating operators involving leptons (such as the lepton EDMs themselves) do not lead to large threshold corrections to $\bar \theta$. We discuss paramagnetic systems further at the end of this section. Instead, we are lead to consider the correlations between hadronic EDM predictions. In Fig.~\ref{EDMcorrelation} we show contours consistent with a pure-$\bar \theta$ scenario for pairs of hypothetical hadronic EDM observations. As all EDMs depend on a single parameter, $\bar \theta$, all EDMs are linearly correlated, but the current theoretical uncertainties lead to contours and regions instead of lines. 

The left panel of Fig.~\ref{EDMcorrelation} shows that if $\bar \theta$ is the only source of CP violation in these systems, the diamagnetic EDMs are expected to be small with respect to the neutron EDM due to  Schiff screening (note that $d_{\mathrm{Hg}}$ has been multiplied by a factor $100$ to make the contour visible). The associated uncertainties in these EDMs are also large enough that given a measurement of the neutron EDM, the sign of $d_{\mathrm{Hg}}$ and $d_{\mathrm{Ra}}$ cannot be predicted. Nevertheless, ratios of $|d_{\mathrm{Ra}}/d_n| \gtrsim 1$ and  $|d_{\mathrm{Hg}}/d_n| \gtrsim 5\cdot 10^{-3}$ would point towards dimension-six sources of CP violation. 

How do the necessary sources compare with the operators listed in Table~\ref{fig:optable}, producing quadratic divergences in $\bar\theta$? The set of dimension-six operators relevant for hadronic and nuclear CP violation was derived in Ref.~\cite{deVries:2012ab}, starting from the SMEFT operators and matching to a low-energy EFT around 2 GeV. At this scale, the non-leptonic operators that induce hadronic and nuclear EDMs include quark EDMs, quark chromo-EDMs, the Weinberg operator, and several four-quark operators. The former three are directly induced by the operators $\mathcal O_{dB}, \mathcal O_{dW}, \mathcal O_{uB}, \mathcal O_{uW}, \mathcal O_{dG}, \mathcal O_{uG},  \mathcal O_{\tilde G}$, all of which appear in Table~\ref{fig:optable}. The four-quark operators can be divided in two sets. The first set consists of operators induced by $\mathcal O_{quqd(1)}$ and $\mathcal O_{quqd(8)}$, which also appear in Table \ref{fig:optable}. The final two four-quark operators do not appear in Table \ref{fig:optable} as they are not $SU_L(2)$ gauge invariant. However, they are induced after electroweak symmetry breaking via a combination of SM weak interactions and $\mathcal O_{Hud}$, which does appear in Table~\ref{fig:optable}. Thus, deviations from the predictions of the pure $\bar \theta$ scenario in low-energy EDM measurements can be explained by the same SMEFT operators that point to large threshold corrections to $\bar\theta$.

A small aside is in order here. In principle, the Hg EDM, being an atomic system, gets contributions from the electron EDM and semi-leptonic electron-nucleon operators. The electron EDM  clearly does not imply large corrections to $\bar \theta$. However, a nonzero Hg EDM in upcoming experiments would imply values of $d_e$ that are already ruled out by paramagnetic EDM experiments. The most relevant semi-leptonic electron-nucleon interaction that could induce $d_{\mathrm{Hg}}$ is the tensor operator $\bar e i\sigma^{\mu\nu}\gamma^5 e\,\bar N \sigma_{\mu\nu} N $, which is mainly induced by the SMEFT operator $\epsilon^{ef} \overline{L^e_{Li}} \sigma^{\mu\nu} e_{Rj} \overline{Q^f_{Lk}} \sigma_{\mu\nu}  u_{Rl}$. While this operator does not appear in Table~\ref{fig:optable}, it mixes under one-loop RGE with $\op_{lequ}^{(1)}$, which does appear in the table.  As such, values of $|d_{\mathrm{Hg}}/d_n| > 5\cdot 10^{-3}$ indeed imply dimension-six operators that induce $\bar \theta$ threshold corrections.

The right panel shows similar contours, but  for EDMs of the light ions (proton, deuteron, and helion.) Since the theoretical control is typically better, we see clearly that $d_p$ and $d_n$ are anti-correlated, while $d_{\mathrm{He}}$ and $d_n$ are correlated, and $1<d_{\mathrm{He}}/d_n < 2.5$.  The deuteron EDM depends on the sum of nucleon EDMs and on $\bar g_1$, both of which are poorly known in terms of $\bar \theta$. As such, we cannot predict the sign of $d_D$ even if $d_n$ is known. Nevertheless we still expect $|d_D| \leq |d_n|$. Again, for many BSM sources of CP violation these predictions can be quite different. In models where quark EDMs are the dominant source of CP violation (for instance in split-SUSY models \cite{Giudice:2005rz}), the neutron and proton EDM are expected to be of similar size but the relative sign can be both negative and positive. In those models, we expect $d_{\mathrm{He}} \simeq 0.9 d_n$ in contrast to the $\bar \theta$ predictions. In models with large CP-violating four-quark operators or chromo-EDMs, the EDMs of the deuteron and helion are expected to be significantly larger than the single-nucleon EDMs due to the contributions from the CP-violating nuclear force induced by $\bar g_{0,1}$ \cite{jordyunraveling}.

For completeness, we briefly discuss paramagnetic systems. A nonzero EDM of a paramagnetic system would rule out a pure $\bar \theta$ scenario; however, it does not automatically rule out UV solutions to strong CP. EDMs of systems such as the Tl atom and ThO and HfF polar molecules are essentially dominated by two CP-violating effective interactions: the electron EDM and scalar electron-nucleon interactions. If the electron EDM is dominant, there is not necessarily a large threshold correction to $\bar\theta$. On the other hand, the semi-leptonic scalar electron-nucleon interaction arises from the $\op_{lequ}^{(1)}$ operator, which does generate a quadratic divergence. Recent studies \cite{Chupp:2014gka,Chupp:2017rkp, Fleig:2018bsf} have shown that measurements of several paramagnetic systems, in addition to the diamagnetic Hg EDM, can isolate the dominant source of CP violation (i.e. the electron EDM or the scalar electron-nucleon coupling). As such, even paramagnetic EDMs can in some cases provide a useful probe of radiative corrections to $\bar\theta$.

\section{Collider probes}
\label{sec:collider}

Collider experiments are sensitive to a wide range of signatures associated with SMEFT operators. In our context, an advantage of colliders is that they provide more direct access to the individual dimension--6 CPV operators that renormalize $\overline{\theta}$ than EDM searches. On the other hand, colliders can only probe a limited subset of the operators in Table~\ref{fig:optable}, due to large backgrounds associated with light quark/gluon jets, and the challenge of constructing measurable CP-sensitive observables. Below, for illustration, we will restrict our attention to processes involving top quarks. Tops have long been recognized as offering especially promising tests of CP-violation beyond the Standard Model~\cite{Kane:1991bg,Atwood:1991ka, Bernreuther:1992be, Atwood:2000tu}, since cross sections are large and tops can be efficiently identified and reconstructed at colliders.\footnote{The CP-odd operator $O_{H\tilde G}$ can also be probed at hadron colliders, for example, via angular correlations in $h+jj$ events~\cite{Dolan:2014upa}. Recently, Ref.~\cite{Bernlochner:2018opw} performed an analysis of this channel at the LHC and reported a  CP-odd asymmetry in $\Delta\phi_{jj}$ of $0.3\pm0.2$. Assuming the significance grows in the future, Ref.~\cite{Bernlochner:2018opw} concluded that values of $|c_{H\tilde G}/\Lambda^2| \simeq 0.1\, \mathrm{TeV}^{-2}$ can reproduce the central value. However, a study of hadronic EDMs induced by $O_{H\tilde G}$ concluded that  $|c_{H\tilde G}/\Lambda^2| < (7 \cdot 10^{-3})\, \mathrm{TeV}^{-2}$ from the neutron EDM limit using conservative values of the relevant matrix elements~\cite{Chien:2015xha}. Limits on operators with Higgs fields and electroweak field strengths are even stronger because they induce the electron EDM at one loop. While EDM limits can be avoided by cancellations with other contributions, significant fine tuning (at the few-percent level) is required to align the EDM limits with the present hint of a CP-odd asymmetry.}

Several top quark operators appearing in Table~\ref{fig:optable}, including\footnote{We define the third-generation operators $\mathcal{O}_{t X} \equiv \widetilde{\mathcal{O}}_{uX}^{33}$, where $\widetilde{\mathcal{O}}_{uX}^{ij}$ represents the operator $\mathcal{O}_{uX}^{i^\prime j^\prime}$ in Table~\ref{fig:optable} rotated into the quark mass basis. Here $i,j$ and $i^\prime, j^\prime$ are generation indices in the mass and gauge eigenstate bases, respectively. The corresponding Wilson coefficients $c_{tX}$ are defined analogously.} $\mathcal{O}_{tH}$, $\mathcal{O}_{tB}$, and $\mathcal{O}_{tW}$, contribute to the electron EDM at two loops, and are thus quite constrained by limits on $d_e$~\cite{Brod:2013cka, Chien:2015xha, Cirigliano:2016nyn, Fuyuto:2017xup}, barring cancellations between contributions. In what follows we focus on the chromo-dipole moment operator $\mathcal{O}_{tG}$, since it does not contribute to the $e$-EDM at one or two loops. It can have a sizable impact on the neutron EDM~\cite{Kamenik:2011dk,Chien:2015xha, Cirigliano:2016nyn}, but due to the weaker limits and significantly larger uncertainties than those associated with the $e$-EDM, it is possible for $\mathcal{O}_{tG}$ to produce observably large effects at colliders while remaining consistent with EDM bounds within their respective uncertainties.

\subsection{Top quark CEDM operator and EDM constraints}

We consider the top quark chromo-magnetic and chromo-electric dipole moment operators induced by $\mathcal{O}_{tG}$ after electroweak symmetry breaking:
\beq
\mathcal{L} \supset - g_s \frac{\widetilde \mu_t}{2 m_t} \bar{t} \sigma^{\mu \nu}T^a t G_{\mu \nu}^a -  i g_s \frac{\widetilde{d}_t}{2m_t} \bar{t} \sigma^{\mu \nu}T^a \gamma_5 t G_{\mu \nu}^a.
\eeq
 The chromo-magnetic and chromo-electric dipole moments, $\widetilde \mu_t$ and $\widetilde d_t$, can be straightforwardly related to the real and imaginary parts of the corresponding Wilson coefficients before electroweak symmetry breaking: 
\beq 
\widetilde \mu_t = -\frac{2\operatorname{Re}\left(c_{tG}\right) m_t^2}{y_t g_s \Lambda^2}, \quad \widetilde d_t = -\frac{2\operatorname{Im}\left(c_{tG}\right) m_t^2}{y_t g_s \Lambda^2}. 
\eeq
Evidence for a nonzero $\widetilde d_t$ directly implies non-zero components of $\mathcal{O}_{uG}$, and thus large threshold corrections to $\bar\theta$ via Eq.~(\ref{eq:thetacorr}).

Due to the significant hadronic and nuclear uncertainties involved, placing robust limits on $\widetilde{d}_t$ requires some care. Refs.~\cite{Chien:2015xha, Cirigliano:2016nyn} performed an analysis of the constraints on $\widetilde d_t$ from experimental limits on $d_e$, $d_n$, and $d_{\rm Hg}$ using state-of-the art matrix elements, finding an upper bound of
\beq \label{eq:dn_bound}
|\widetilde d_t | \lesssim 2 \times 10^{-2} \quad (90\% \, {\rm C.L., \, current}).
\eeq
In this constraint, $\widetilde d_t$ is evaluated at the scale\footnote{Varying the scale between 1-100 TeV has an $\mathcal{O}(1)$ impact on the bounds, since it impacts the running of the couplings logarithmically. The bound on $\widetilde d_t$ becomes weaker for $\Lambda > 1$ TeV.} $\Lambda =1$ TeV and assuming that only the top CEDM operator is present at these energies. If one instead allows for the presence of other operators with comparable Wilson coefficients, the bound is weakened due to possible cancellations between contributions to the EDMs. The upper limit in Eq.~(\ref{eq:dn_bound}) accounts for the various experimental uncertainties, as well as theoretical uncertainties in the predicted values of $d_n$ and $d_{\rm Hg}$ by varying the relevant hadronic and nuclear matrix elements across their allowed ranges. This bound therefore represents a conservative upper limit on $\widetilde{d}_t$, allowing for possible cancellations between contributions to $d_n$ and/or $d_{\rm Hg}$. As a result, it is significantly weaker than results appearing elsewhere in the literature where all of the uncertainties are not accounted for in this way. If one instead adopts the central values for all matrix elements, one arrives at a significantly more stringent bound, $|\widetilde d_t | \lesssim 1.5 \times 10^{-4}$~\cite{Cirigliano:2016nyn}. This dramatic difference indicates that improvements in the theoretical modeling of the neutron EDM can have a large impact on the allowed values of $\widetilde d_t$. Ref.~\cite{Chien:2015xha} estimates that a robust upper limit analogous to Eq.~(\ref{eq:dn_bound}) of 
\beq 
|\widetilde d_t | \lesssim 8 \times 10^{-4} \quad (90\% \, {\rm C.L., \, improved \, matrix \, elements})
\eeq
can be achieved with realistic improvements in the hadronic and nuclear matrix element uncertainties. Of course more sensitive measurements will also impact these limits. 

\subsection{The top CEDM and CP-sensitive collider observables}

Given the large uncertainties in the EDM bounds, we conservatively adopt Eq.~(\ref{eq:dn_bound}) and investigate the extent to which hadron colliders can directly probe $\widetilde d_t$ at this level and below. Both $\widetilde d_t$ and $\widetilde \mu_t$ impact various CP-insensitive observables at colliders, such as the Higgs and $t \bar{t}$ production rates. While many previous studies have investigated these effects~\cite{Kamenik:2011dk, Hayreter:2013kba, Aguilar-Saavedra:2014iga, Chien:2015xha, Cirigliano:2016nyn}, we instead focus on CP-odd observables sensitive to $\widetilde d_t$ in the dimension--6 SMEFT, as they can provide direct evidence for a large threshold correction to $\bar \theta$. 

To this end, we follow Refs.~\cite{Bernreuther:1998qv,Bernreuther:2010ny, Bernreuther:2013aga, Bernreuther:2015yna, Aaboud:2016bit} and consider CP-odd triple product observables in dileptonic $t \bar{t}$ production at hadron colliders (see also Ref.~\cite{Ma:2017vve} for a study of related observables).  Refs.~\cite{Bernreuther:2013aga, Bernreuther:2015yna} showed that the expectation value of the quantity 
\beq
\mathcal{O}_{\rm CP} \equiv \left(\hat{\ell}_+ \times \hat{\ell}_-\right) \cdot \hat{\mathbf{k}}
\eeq 
is directly related to $\widetilde{d}_t$. Here $\hat{\ell}_+$ and $\hat{\ell}_-$ are the directions of flight of the $\bar \ell$ and $\ell$ in the $\bar t$ and $t$ rest frames, respectively, and $\hat{\mathbf{k}}$ is the $t$ direction of flight in the $t\bar t$ center-of-mass frame. A non-zero $\langle \mathcal{O}_{\rm CP}  \rangle$ results in a non-vanishing CP-asymmetry, $A_{\rm CP}$, defined as
\beq \label{eq:ACP}
A_{\rm CP} \equiv \frac{ N(\mathcal{O}_{\rm CP} > 0) -  N(\mathcal{O}_{\rm CP} < 0)}{N(\mathcal{O}_{\rm CP} > 0) +  N(\mathcal{O}_{\rm CP} < 0)}
\eeq
where $N$ denotes the corresponding number of dileptonic $t\bar t$ events. Standard Model contributions to $\langle \mathcal{O}_{\rm CP}  \rangle$ are negligible, and in the operator basis used here, $\langle \mathcal{O}_{\rm CP}  \rangle$  receives a contribution only from the top chromo-EDM at leading order. Observation of $A_{\rm CP}\neq 0 $ at the LHC or a future collider would imply the need for a low-energy solution to the strong CP problem. 

We extend the results of Refs.~\cite{Bernreuther:2013aga, Bernreuther:2015yna} by estimating the expected sensitivity to $A_{\rm CP}$, and thus $\widetilde{d}_t$, at the high-luminosity  LHC (HL-LHC) with $\sqrt{s}=14$ TeV, as well as a high-energy phase of the LHC (HE-LHC) with $\sqrt{s}=27$ TeV and a future $\sqrt{s}=100$ TeV collider. We account for showering/hadronization and detector resolution effects, which impact the reconstruction of the $t\bar t$ system necessary to determine $\mathcal{O}_{\rm CP}$. For each center of mass energy, we used \texttt{Madgraph 5}~\cite{Alwall:2014hca} to generate $p p \to t\bar t \to b \ell \bar \nu \bar b \bar \ell^\prime \nu^\prime$ Monte Carlo events for various values of $\widetilde{d}_t$, utilizing a model file built by the \texttt{FeynRules} package~\cite{Alloul:2013bka}. Events were then passed to \texttt{Pythia 6}~\cite{Sjostrand:2006za} for showering/hadronization and to \texttt{DELPHES 3}~\cite{deFavereau:2013fsa} for fast detector simulation. For $\sqrt{s} = 14,$ 27 TeV we use the default CMS \texttt{DELPHES} card with the lepton isolation criterion
\beq
\sum_i \frac{p_T^i}{p_T^\ell} < 0.1
\eeq
for both muons and electrons (here $p_T^\ell$ denotes the transverse momentum of the lepton or anti-lepton and $i$ denotes all other particle flow objects within a $\Delta R< 0.5$ cone of $\ell$ and with $p_T^i > 0.1$ GeV). For $\sqrt{s}= 100$ TeV we use the default FCC-hh detector card included in the \texttt{Delphes 3} distribution.

We select events with exactly two identified oppositely-charged leptons with $p_T > 10$ GeV and two $b$-tagged jets, all with $|\eta| < 2.5$ and $p_T > 10$ GeV. To compute $\mathcal{O}_{\rm CP}$ for a given event, we need to reconstruct the $t \bar t$ system. This is non-trivial due to the two neutrinos in the final state. To do so, we take the following simple approach: we determine the neutrino four-momenta by requiring that the $\ell$ and $\bar \nu$ momenta reconstruct to $m_W$, and that the corresponding reconstructed $W$ boson and one of the $b$-jets reconstruct to the top mass, $m_t \approx 172$ GeV. The same is required for the $\bar \ell$ and $\nu$ momenta and the other $b$-jet. If there are multiple real solutions for a given pairing of the $b$-jets with $W$ momenta, we choose the solution minimizing the scalar sum of the neutrino four-momenta,
\beq
\sum_{i=1,2} E^{\nu_i} + \left|p_x^{\nu_i}\right|+\left|p_y^{\nu_i}\right| + \left|p_z^{\nu_i}\right|.
\eeq
 In some cases, both possible pairings of $b$-jets with the reconstructed $W$ bosons yield real solutions to the equations, in which case we select the pairing minimizing the sum of $\Delta R$ values between the reconstructed $W$s and corresponding $b$-jets.  We solve the corresponding system of equations numerically, and obtain a reconstruction efficiency of roughly $50 -70\%$, depending on which numerical solver and algorithm is used. For comparison, the LHC collaborations are able to obtain up to $\sim 90\%$ reconstruction efficiencies using more sophisticated techniques (see e.g.~\cite{Aaboud:2016bit}). Thus, we expect that our sensitivity projections will be conservative from this standpoint.

\begin{figure}[t!]
\begin{center}
\includegraphics[width=0.6\linewidth]{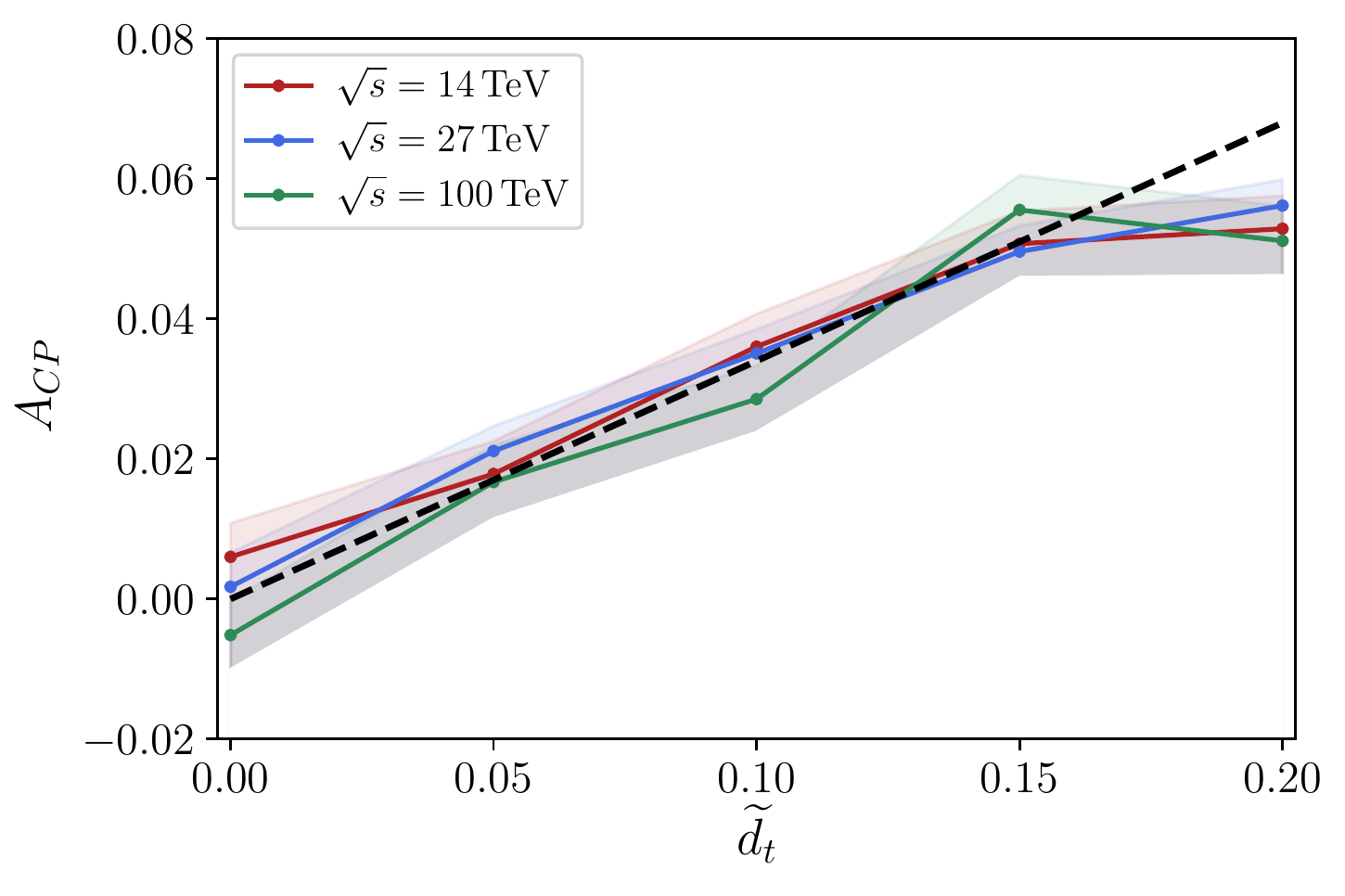}
\caption{Dependence of $A_{\rm CP}$ on $\widetilde{d}_t$ for various collider center-of-mass energies after pre-selection, reconstruction, and requiring $m_{t \bar t} < 1$ TeV. Shaded regions correspond to 1$\sigma$ Monte Carlo uncertainties. Also shown is the heuristic relation $A_{\rm CP} = 0.34 \, \widetilde{d}_t$, which fits the simulated points well for $|\widetilde{d}_t|\lesssim 0.15$, where the linear approximation begins to break down. } 
\label{fig:A_vs_dt}
\end{center}
\end{figure} 

With the $t\bar t$ system reconstructed, we compute $A_{\rm CP}$ for each event sample. As discussed in Ref.~\cite{Bernreuther:2015yna}, for small enough values of $|\widetilde{d}_t|$, $A_{\rm CP} \propto \widetilde{d_t}$, since it is dominated by the interference piece between the CEDM operator and SM contribution to the $t\bar t$ production cross-section. We show the dependence of $A_{\rm CP}$ on $\widetilde{d}_t$ after our preselection cuts and reconstruction in Fig.~\ref{fig:A_vs_dt} for the HL-LHC, HE-LHC, and 100 TeV collider. Also shown are the approximate 1$\sigma$ Monte Carlo uncertainties reflecting the limited number of events generated for each point. The linear behavior is clear for all three collider energies for $\widetilde{d}_t \lesssim 0.15$. The dashed black line in Fig.~\ref{fig:A_vs_dt} corresponds to the relation
\beq \label{eq:ACP_dt}
A_{\rm CP} \simeq 0.34 \, \widetilde{d}_t
\eeq
which we find to be a good match to our Monte Carlo results.

For the relatively wide range of collider center-of-mass energies we consider, it is important to ensure the validity of the EFT approach in our analysis. For all values of $\sqrt{s}$ considered, the $\widetilde{d}_t$ contribution to the total $t\bar t$ cross-section is less than $50\%$ of the SM contribution for $|\widetilde{d}_t|\lesssim 0.15$ where the linear approximation for $A_{\rm CP}$ holds. This suggests that the effect of the dimension--6 operators on $t\bar t$ production is perturbative for the momentum scales relevant for our analysis and that corrections from higher-dimension operators should be under control. Furthermore, we require the $t \bar t$ invariant mass to satisfy $m_{t \bar t}< 1$ TeV throughout our analysis. In particular, this requirement is reflected in the results of Fig.~\ref{fig:A_vs_dt}. Our results are rather insensitive to the $m_{t\bar t}$ cut, signaling that the effects of $\widetilde d_t$ on $A_{\rm CP}$ are dominated by events with sub-TeV momentum transfer and  safely in the domain of validity of the EFT. 

Our analysis neglects the effects of backgrounds mimicking dileptonic $t\bar t$ events. In the Standard Model, none of these processes contribute appreciably to the numerator of Eq.~(\ref{eq:ACP}), but they would contribute to the denominator, and thus somewhat weaken the projected sensitivity. However, since we expect genuine $t\bar t$ events to strongly dominate the denominator, our projections should not be significantly affected by the inclusion of these backgrounds.

\subsection{Results}

Using the relation between $A_{\rm CP}$ and $\widetilde{d}_t$ in Eq.~(\ref{eq:ACP_dt}), we can estimate the  sensitivity of the various colliders to $\widetilde{d}_t$ and compare to constraints from EDM experiments. Given the Standard Model hypothesis, a $\sim 1 \sigma$ statistical fluctuation in the observed value of $A_{\rm CP}$ would correspond to
\beq
\Delta A_{\rm CP} \simeq \left(\left(\sigma \times {\rm BR}\right) \times \int \mathcal{L} \times \left(A\times \varepsilon \times \varepsilon_{\rm reco} \right)\right)^{-1/2}
\eeq 
assuming a large number of $t \bar t$ events so that Gaussian statistics are appropriate and that the SM contribution dominates the dileptonic $t\bar t$ cross-section,  $\sigma \times {\rm BR}$. Here, $\int \mathcal{L}$ is the total integrated luminosity and $\varepsilon_{\rm reco}$ is the efficiency for reconstructing the $t \bar t$ system, which we take to be $\simeq 70\%$. $A\times \varepsilon$ is the acceptance $\times$ efficiency for identifying two oppositely charged leptons and two $b$-tagged jets meeting the kinematic requirements above at a given collider. We find $A\times \varepsilon  \simeq 8\%$ for our HL-LHC and HE-LHC analyses, while for the 100 TeV case we find $A\times \varepsilon \simeq 15\%$, reflecting the higher identification and tagging efficiencies in the FCC-hh \texttt{DELPHES} card. Requiring $A_{\rm CP} > 5 \times \Delta A_{\rm CP}$, and using Eq.~(\ref{eq:ACP_dt}), we obtain $\sim 5 \sigma$ sensitivity projections for $\widetilde{d}_t$. We find that sensitivity to
\beq
\begin{aligned}
|\widetilde{d}_t| \, \gtrsim \, & \, 6.9 \times10^{-3} \quad ({\rm HL -LHC})\\
&\, 3.8 \times10^{-3} \quad ({\rm HE -LHC})\\
&\, 8.3 \times10^{-4} \quad ({\rm 100 \, TeV})\\
\end{aligned}
\eeq
can be reached assuming $\int \mathcal{L} = 3$ ab$^{-1}$. The reach of course improves with increased efficiencies and integrated luminosity. With $A\times \varepsilon \times \varepsilon_{\rm reco} \simeq 20\%$ and $\int \mathcal{L} = 30$ ab$^{-1}$, for example, our analysis suggests that a 100 TeV collider could probe $|\widetilde{d}_t|\gtrsim 1.9 \times 10^{-4}$. The sensitivities above only reflect statistical uncertainties; future work (and detector designs for 27 and 100 TeV) will be required to sharpen the above estimates by including the effects of systematic uncertainties. Our results are therefore optimistic from this standpoint. 

Comparing these results with the EDM constraints on $\widetilde{d}_t$, we see that all three colliders studied above could observe a non-zero $A_{\rm CP}$ at the $\sim 5 \sigma$ level while remaining consistent with current EDM bounds, provided one adopts a conservative interpretation of the various uncertainties in hadronic and nuclear matrix elements. If central values are adopted, a 100 TeV collider could still access the allowed region, provided that the neutron EDM bounds do not significantly tighten before then. In any case, cancellations between various operators could in principle allow for the HL-LHC, HE-LHC, or a 100 TeV collider to discover a non-zero $A_{\rm CP}$, and hence large threshold corrections to $\bar \theta$ while remaining consistent with improved EDM limits.  Using Eq.~(\ref{eq:ACP_dt}), new CPV physics at scales of order 3, 5, and 10 TeV  can be probed by the HL-LHC, HE-LHC, and a future 100 TeV collider, respectively, assuming a Wilson coefficient $c_{tG}\sim \mathcal{O}(1)$.

\section{Summary}
\label{sec:summary}

Searches for new sources of CP violation beyond the Standard Model are of fundamental importance, probing  symmetry structure and the origin of the matter-antimatter asymmetry. We have argued that they can play an additional valuable role in discriminating how nature solves the strong CP problem. Even if new physics is too heavy to be produced on-shell, if signatures of a broad class of dimension-6 operators are experimentally observed, it will strongly disfavor models in which $\theta=0$ is an ultraviolet boundary condition and provide indirect support for the existence of a QCD axion.

Both low energy and high energy experiments are sensitive to these operators. At low energies, correlations among two or more hadronic EDM measurements can be used to reject a pure-$\theta$ explanation over a wide range of parameter space, limited primarily by theoretical uncertainties. High energy colliders can also access new CP-violating operators. As an example, we have analyzed the HL-LHC, HE-LHC, and 100 TeV collider reach for a nonzero top chromo-EDM in angular observables. Both classes of experiments are complementary: EDM experiments offer high sensitivity, while colliders provide more direct access to individual operators, particularly in the third generation, and are insensitive to hadronic uncertainties. The insight such discoveries could provide into the resolution of the strong CP problem further increases the value of these searches.

\vskip 1cm
\noindent
{\bf Acknowledgements:}  We thank Hao-Lin Li, Michael Ramsey-Musolf, Adam Ritz and Felix Yu for discussions. The work of PD and JK was supported by NSF grant PHY-1719642. JK and DS gratefully acknowledge the hospitality of the Aspen Center for Physics, supported by National Science Foundation grant PHY-1607611, where a portion of this work was completed. The work of KF was supported by Department of Energy grant DE-SC0011095. The work of DS was supported in part by Department of Energy grant DE-SC0014129.

\appendix
\section*{Appendix}
Here we collect conventions used in the calculation of quadratic divergences in Sec.~\ref{sec:quad}. 
The Standard Model Lagrangian is given by
\begin{align*}
\lag_\text{SM} =& \sum_{F=B,W,G} -\frac{1}{4} F_{\mu\nu} F^{\mu\nu} + \sum_{\psi=Q_L,L_L,u_R,d_R,e_R} i \overline{\psi^i} D \psi^i + |D_\mu H|^2 - V(|H|^2) \nonumber \\
&+ \frac{\theta g_s^2}{16 \pi^2} G^a_{\mu\nu} \tilde G^{a \mu\nu}
-( Y_u^{ij} \overline{Q_{Li}} \tilde{H} u_{Rj}
+ Y_d^{ij} \overline{Q_{Li}} H d_{Rj}
+ Y_e^{ij} \overline{L_{Li}} H e_{Rj} + \text{h.c.} ) .
\end{align*}
The conventions implicit in $\lag_\text{SM}$ align with those of \cite{Grzadkowski:2010es}. To wit, we use four component spinors for the matter fields, subscripts $L$ and $R$ denoting the action of the projection operator $P_{{\overset{\scriptscriptstyle L}{ \scriptscriptstyle R}}} = \frac{1}{2} (1 \mp \gamma^5)$. $\tilde G^{\mu\nu} = \frac{1}{2} \epsilon^{\mu\nu\alpha\beta} G_{\alpha \beta}$, where $\epsilon^{\mu\nu\alpha\beta} = -\frac{1}{4} i \Tr(\gamma^\mu \gamma^\nu \gamma^\alpha \gamma^\beta \gamma^5)$. The gauge fields are normalized such that the covariant derivative $D_\mu Q_{Li} = (\partial_\mu + \frac{1}{6} i g^\prime B_\mu + \frac{1}{2} i g W^i_\mu \tau^i + i g_s G^a_\mu T^a) Q_{Li}$, where $\Tr(T^a T^b) = \frac{1}{2} \delta^{ab}$ and $\Tr(\tau^i \tau^j) = 2 \delta^{ij}$, and $\{B_\mu,W^i_\mu,G^a_\mu\}$ are the vector potentials feeding into the field strengths via $G^a_{\mu\nu} \equiv \partial_\mu G^a_\nu - \partial_\nu G^a_\nu - g_s f^{abc} G^b_\mu G^c_\nu$ and so forth. $\overline{Q_L} \tilde H \equiv \epsilon^{ab} \overline{Q_L}^a H^b$ in terms of weak isospin indices $a,b$, and $\epsilon^{12}=+1$.

\bibliography{theta_smeft}{}
\bibliographystyle{utphys}

\end{document}